\documentclass[aps,twocolumn,gbroupedaddress,reprint,amsmath,amssymb]{revtex4-2}
\usepackage[utf8]{inputenc}
\usepackage[T1]{fontenc}
\usepackage{amsmath}
\usepackage{amsfonts}
\usepackage{amssymb}
\usepackage{graphics,graphicx}
\usepackage{graphicx}
\usepackage{subfigure} 
\usepackage{wrapfig} 
\usepackage{epstopdf}
\usepackage{xcolor}
\graphicspath{{figuras/}}

\usepackage{ulem}
\usepackage[breaklinks=true]{hyperref}
\usepackage{setspace}
\usepackage{graphicx}
\usepackage{color}

\begin{document}

\title{Novel five-dimensional rotating Lifshitz black holes with electric and axionic charges.}
\author{Mois\'es Bravo-Gaete}
\email{mbravo@ucm.cl, moisesbravog@gmail.com}
\affiliation{Departamento de Matem\'atica, F\'isica y Estad\'istica, Facultad de Ciencias
B\'asicas, Universidad Cat\'olica del Maule, Casilla 617, Talca, Chile.}

\author{Jhony A. Herrera-Mendoza}
\email{jhonyahm@gmail.com}
\affiliation{Escuela de F\'isica, Facultad de Ciencias, Universidad Nacional Aut\'onoma de Honduras,
Blvr. Suyapa, Tegucigalpa, Municipio del Distrito Central 11101, Honduras.}
\affiliation{Instituto de Física, Benem\'erita Universidad Aut\'onoma de Puebla, Edificio IF-1, Ciudad Universitaria, Puebla, Pue. 72570, M\'exico.}

\author{Julio Oliva}
\email{julioolivazapata@gmail.com} \affiliation{Departamento de F\'isica, Universidad de Concepci\'on,
Casilla, 160-C, Concepci\'on, Chile.}

\author{Xiangdong Zhang}
\email{scxdzhang@scut.edu.cn}
\affiliation{School of Physics and Optoelectronics, South China University of Technology, Guangzhou 510641, China.}

\date{\today}          
\begin{abstract}
\textcolor{black}{In this work, we construct a new family of exact five-dimensional charged and rotating asymptotically Lifshitz black holes}. The spacetime solves Einstein equations coupled to a dilaton, two Abelian gauge fields, and axionic scalars supplemented by two generalized Chern-Simons terms. This configuration is characterized by a range of the free dynamical exponent $z$ and possesses nontrivial thermodynamical parameters, where we verify the first law of black hole thermodynamics and derive the corresponding Smarr relation. \textcolor{black}{As an application of this new gravitational background}, we then investigate a holographic superconductor in the rotating Lifshitz background. We study the condensation of the scalar operator and the AC conductivity of the dual system. These results show that increasing the rotation parameter suppresses the condensate and weakens the superconducting phase, while increasing the dynamical critical exponent enhances the superconducting order. \textcolor{black}{To the best of our knowledge, these solutions provide the first explicit
family of five-dimensional rotating Lifshitz black holes supported
simultaneously by electric and axionic charges.}
\end{abstract}

\maketitle
\section{Introduction}\label{intro}
The holographic dictionary, originally conceived in \cite{tHooft:1993dmi,Susskind:1994vu}, found a precise formulation in the Maldacena Conjecture \cite{Maldacena:1997re}, according to which string theory on AdS$_5\times S^5$ is dual {to $\mathcal{N}=4$ Super-Yang-Mills theory with gauge group $SU(N)$}. This UV-complete formulation of holography has been useful {for exploring} many aspects of both sides of the duality in non-perturbative regimes. In the more general framework of holography \cite{Witten:1998qj,Gubser:1998bc}, gravitational physics in {anti-de Sitter (AdS)} in $D-$dimensions can be used to describe strongly coupled systems in one dimension less, which can be realized at finite temperature if the bulk contains a black hole. \textcolor{black}{Strongly coupled systems are ubiquitous in condensed matter physics, and
high-temperature superconductors provide an important example where the
duality can be applied.} In this direction, pioneering works in holographic superconductivity have captured key features of high-temperature superconductors by considering Abelian-Higgs-like models in Schwarzschild {AdS} black hole backgrounds \cite{Gubser:2008px, Hartnoll:2008vx, Nakano:2008xc, Hartnoll:2008kx}.
Subsequently, the inclusion of an external magnetic field to the superconductor allowed the holographic realization of the London equation and the Abrikosov vortex lattice \cite{Montull:2009fe, Maeda:2009vf, Domenech:2010nf, Montull:2011im, Montull:2012fy, Salvio:2012at, Salvio:2013jia, Donos:2020viz,  Xia:2021jzh, delaCruz-Lopez:2024chw}. Further extensions of holographic superconductivity have explored non-commutative AdS backgrounds, where the effects of non-commutativity are found to modify the condensation properties \cite{delaCruz-Lopez:2024ync}.

Many extensions of these ideas have been {developed over the last decades. Of particular relevance for the present work is non-relativistic holographic duality,} according to which gravitational physics on spacetimes with an asymptotic anisotropic scaling symmetry $t\rightarrow \lambda^{z}t$, $r\rightarrow \lambda^{-1} r$  and $\vec{x}\rightarrow\lambda\vec{x}$, is dual to strongly coupled non-relativistic systems. \textcolor{black}{This symmetry is not sufficient to completely fix the form of the
two-point functions in the dual theory. Nevertheless, it characterizes
a non-relativistic system with dynamical exponent $z$ at a Lifshitz
fixed point. This setup can also be considered at finite temperature } and, in recent years, the study of black holes with Lifshitz asymptotic structure has gained considerable interest within the non-relativistic domains of the gauge/gravity duality \cite{Kachru:2008yh}. In this framework, holographic superconductors constructed in Lifshitz backgrounds have shown that the critical temperature, order parameter, and critical magnetic field depend sensitively on the value of the dynamical critical exponent, thereby revealing a richer structure {compared to} their AdS counterparts \cite{Taylor:2008tg,Bu:2012zzb, Lu:2013tza, Zhao:2013pva, Herrera-Mendoza:2022whz}. 

In recent works, the inclusion of rotating black hole backgrounds in the construction of holographic superconductors has uncovered {intriguing} connections between the black hole rotation and the suppression of the superconducting state \cite{Srivastav:2019ixc, Lin:2014tza, Herrera-Mendoza:2024vfj, Bravo-Gaete:2025vyd}. Consequently, the study of holographic superconductivity using rotating Lifshitz black hole backgrounds emerges as a natural generalization, \textcolor{black}{allowing for a deeper understanding} of how anisotropic scaling and background rotation interplay in the formation of the superconducting state.

\textcolor{black}{To the best of our knowledge, the solutions presented in this work
constitute the first explicit family of five-dimensional rotating
asymptotically Lifshitz black holes supported simultaneously by
electric and axionic charges. The construction of exact rotating Lifshitz geometries is known to be
technically challenging, and explicit examples remain relatively rare.
The solutions obtained here, therefore, provide a new gravitational
background that can be used to explore non-relativistic holography
beyond static configurations. These solutions provide a new gravitational arena} to explore non-relativistic gauge/gravity duality in the presence of rotational degrees of freedom. Moreover, the holographic superconductor constructed on top of these backgrounds allows us to disentangle and systematically analyze the competing effects of anisotropic scaling and rotation on the formation of a superconducting phase.

{The main purpose of this work is twofold: (i) On the gravitational side, we construct and analyze a new family of exact rotating Lifshitz black holes and study their thermodynamic properties. On the holographic side, (ii) we use these solutions as backgrounds to build a holographic superconductor and to study the behavior of the condensate as well as the AC conductivity.} \textcolor{black}{In this sense, the holographic-superconductor analysis should be viewed
as an application of the new gravitational background, aimed at
illustrating how rotation and anisotropic scaling affect the formation
of the superconducting phase.}

The plan of the paper is organized as follows: In Sec. \ref{sol} we present the model and construct the rotating Lifshitz black hole solutions, \textcolor{black}{while the thermodynamic analysis of this configuration is
performed in Section \ref{termo}}. In Sec. \ref{hol-sup}, we study the holographic superconductor, analyzing the condensation process and the electric response of the system. Finally, in Sec. \ref{conclusions} we summarize our results and discuss possible future open problems.

\section{Rotating Lifshitz Black Holes in five dimensions }\label{sol}

In this paper, we consider the supergravity-inspired five-dimensional model:

\begin{eqnarray}\label{eq:bgaction}
 S&=&\frac{1}{2 \kappa}  \int d^5x\sqrt{-g}\Bigg[R-2\Lambda\nonumber-\frac{1}{2} \partial_\mu \phi \partial^{\mu} \phi\nonumber\\
 &-&\sum_{i=1}^{2} \Big( e^{\alpha_i \phi} F^{(i)}_{\mu\nu}F^{(i)\mu\nu}+\frac{1}{2} e^{\eta \phi}  \partial_{\mu} \psi^{(i)} \partial^{\mu} \psi^{(i)}\Big)\Bigg]\nonumber\\
 &+&2 \sum_{i=1}^{2}\frac{\lambda_i}{\kappa}\int \Big(A^{(i)}\wedge H^{(i)}\wedge K^{(i)}\Big),
\end{eqnarray}
where $H^{(i)}=dB^{(i)}$ and $K^{(i)}=dC^{(i)}$. \textcolor{black}{The field content, consisting of the graviton $g_{\mu\nu}$, a scalar
field $\phi$, two Abelian gauge fields $A_\mu^{(i)}$ dilatonically
coupled to $\phi$, and the axionic scalars $\psi^{(i)}$, naturally
arises from} dimensional reductions of supergravity (see e.g. \cite{Freedman:2012zz}). Here, the two derivative action is supplemented by  Chern-Simons terms,  constructed out of two, non-dynamical, one-form doublets {$B^{(i)}_\mu$ and $C^{(i)}_\mu$ \cite{Bravo-Gaete:2025lgs}. The coupling constants of the model are $\lambda_{i}$, $\eta$ and $\alpha_{i}$, with $i\in \{1,2\}$.} The field equations are given by:
\begin{eqnarray}
 \mathcal{E}_{\mu \nu}&:=&G_{\mu \nu}+\Lambda g_{\mu \nu}-\left(\frac{1}{2} \partial_\mu \phi \partial_\nu \phi-\frac{1}{4} g_{\mu \nu} (\partial_\sigma \phi \partial^{\sigma}\phi)\right)\nonumber\\
 &-&\sum_{i=1}^{2} \Big[ e^{\alpha_i \phi} \Big( 2  F^{(i)}_{\mu \sigma}F_{\nu}^{(i)\sigma}-\frac{1}{2} g_{\mu \nu} F^{(i)}_{\rho\sigma}F^{(i)\rho\sigma}\Big)\\
 &+&e^{\eta \phi} \Big( \frac{1}{2} \partial_\mu \psi^{(i)} \partial_\nu \psi^{(i)}-\frac{1}{4} g_{\mu \nu} \big(\partial_\sigma \psi^{(i)} \partial^{\sigma}\psi^{(i)}\big) \Big)\Big]\nonumber\\
 &=&0,\nonumber\\
\mathcal{E}_{\psi}^{(i)}&:=&\nabla_{\mu}\left(e^{\eta \phi} \nabla^{\mu} \psi^{(i)}\right)=0, \\
 \mathcal{E}_{\phi}&:=& \Box{\phi}\nonumber\\
 &-&\sum_{i=1}^{2}\Big( \alpha_i e^{\alpha_i \phi} F^{(i)}_{\mu\nu}F^{(i)\mu\nu}+\frac{\eta}{2} e^{\eta \phi} \partial_{\mu} \psi^{(i)} \partial^{\mu} \psi^{(i)}\Big)\\
 &=&0,\nonumber\\
 J^{(i)\nu}&:=&\nabla_\mu \left(e^{\alpha_i \phi} F^{(i)\mu\nu}\right)-\lambda_i *\left(H^{(i)}\wedge K^{(i)}\right)^{\nu}=0,\label{eq:J}
\end{eqnarray}
while the Abelian field strength doublets $H^{(i)}$ and $K^{(i)}$ are constrained by \cite{Deshpande:2024vbn,Hale:2024zvu}:
\begin{equation}\label{eq:constraint}
F^{(i)} \wedge H^{(i)} = 0,\quad F^{(i)} \wedge K^{(i)} = 0, \quad i \in \{1,2\}. 
\end{equation}
\textcolor{black}{We normalize the bare cosmological constant as}
\begin{equation}\label{eq:lambda}
\Lambda=-\frac{(z+3)(z+2)}{2\ell^2}.
\end{equation}
This theory admits the following charged, spinning, asymptotically Lifshitz black hole solution
\begin{eqnarray}\label{eq:metric}
ds^2&=&-N^2(r)f(r) dt^2+\frac{dr^2}{f(r)}+r^2 \big(dx_1+N^{x_1}(r) dt\big)^2\nonumber\\
&+&r^2\sum_{i=2}^3 dx^2_i,
\end{eqnarray}
where $t \in \mathbb{R}$, $r \ge 0$, the rotation direction $x_1$ is an angular coordinate with identification $x_1 \sim x_1 + \sigma_1$, while
$x_2$ and $x_3$ are planar coordinates compactified as $0 \le x_i \le \sigma_i$,  and
\begin{eqnarray}
f(r)&=&\frac{r^2}{\ell^2}-\frac{M \ell^{z+1}}{r^{z+1}}+\frac{\ell^{2(z-1)} (5-z)}{2 (z-2)} \frac{J^2}{r^6}, \nonumber\\ 
N^{x_1}(r)&=&\frac{J}{r^{5-z}}, \qquad N(r)=\left(\frac{r}{\ell}\right)^{z-1},\label{eq:f-N}
\end{eqnarray}
where $M$ and $J$ are integration constants. {The matter fields and coupling constants} $\eta,\ \alpha_1$ and $\alpha_2$ are fixed as:
\begin{eqnarray}
e^{\phi}&=&\mu r^{\sqrt{6(z-1)}},\quad \alpha_1=\eta=-\frac{6}{\sqrt{6(z-1)}},\nonumber\\
\alpha_2&=&\frac{2}{\sqrt{6(z-1)}},\label{eq:phi}\\
A^{(1)}_\mu dx^{\mu}&=& \sqrt{\frac{2(z-1)}{(z+3)}}\frac{\mu^{\frac{3}{\sqrt{6(z-1)}}}}{2 \ell^z}\left(r^{z+3}-r_{h}^{z+3}\right) dt,\,\nonumber\\
A^{(2)}_\mu dx^{\mu}&=& \displaystyle{\sqrt{\frac{3(3-z)}{4 (z-2)}}\frac{J}{\mu^{\frac{1}{\sqrt{6(z-1)}}}}\left(\frac{1}{r^{5-z}}-\frac{1}{r_h^{5-z}}\right) dt,}\nonumber\\
\psi^{(i)}(x_{i+1})&=&(5-z)\ell^{z-1} \mu^{\frac{3}{\sqrt{6(z-1)}}} J\, x_{i+1}, \quad i \in \{1,2\}.\nonumber
\end{eqnarray}
Here, $r_h$ is the location of the event horizon defined as the largest root of $f(r)$ in \eqref{eq:f-N}, {{and $\mu$ is an arbitrary integration constant \footnote{Here we note that, as was shown in \cite{Tarrio:2011de}, the integration constant $\mu$ is fixed to ensure consistency with the desired
Lifshitz asymptotic behavior.}}}. The Lifshitz dynamical exponent is given by $z$, \textcolor{black}{and in our case $z\in (2,3)$ ensures the interpretation of a rotating black hole on a background with anisotropic scaling symmetry. The endpoint $z=3$ requires a separate treatment, and the asymptotic expansions of the gauge fields develop logarithmic terms. For this reason, we restrict our analysis to $z\in(2,3)$.} Notice that $z$ is indeed a free parameter within this range, and is not related to the coupling constants of the theory, neither to other integration constants of the solutions. \textcolor{black}{This is possible only if the field equations $J^{(i)\nu}=0$ (\ref{eq:J}) are
satisfied, which is ensured provided that the one-form doublets
$B^{(i)}$ and $C^{(i)}$ satisfy:}
\begin{eqnarray}
\lambda_1\, *\left(H^{(1)}\wedge K^{(1)}\right)^{x_1}&=&-\frac{\ell^{z-2}\,(5-z)}{6 \mu^{\frac{3}{\sqrt{6(z-1)}}} r^8}\nonumber\\
&\times&\sqrt{18 (z+3) (z-1)}\,J,\label{Jphi1}\\
\lambda_2\, *\left(H^{(2)}\wedge K^{(2)}\right)^{x_1}&=&\frac{\ell^{2(z-1)}\mu^{\frac{1}{\sqrt{6(z-1)}}}}{2\sqrt{z-2}\,r^8}\nonumber\\
&\times&\sqrt{3(3-z)}(5-z)^2\,J^2.\label{Jphi2}
\end{eqnarray}
This is achieved by the following expressions
\begin{eqnarray*}
B^{(i)}_{\mu}dx^{\mu}&=&\left(\frac{t}{\lambda_i}\right) dx_2,\quad i \in \{1,2\}.
\end{eqnarray*}
\begin{eqnarray}
C^{(1)}_{\mu}dx^{\mu}&=&\left(\frac{\sqrt{2 (z+3) (z-1)}\,J}{2 \ell \mu^{\frac{3}{\sqrt{6(z-1)}}} r^{5-z}}\right) dx_3, \label{eq:BC}\\
C^{(2)}_{\mu}dx^{\mu}&=&\left(\frac{\ell^{z-1}\mu^{\frac{1}{\sqrt{6(z-1)}}}\sqrt{3(3-z)}(5-z)\,J^2}{2\sqrt{z-2}\,r^{5-z}}\right) dx_3.\nonumber
\end{eqnarray}

From the line element (\ref{eq:metric}), we note that the scalar curvature $R$ can be expressed as:
\begin{eqnarray*}
R&=&6-\frac{3f'(z+1)}{r}-\frac{2 f (2+z^2)}{r^2}-f''\nonumber\\
&+&\frac{\ell^{2(z-1)}}{2 r^{2z-4}} \big((N^{x_1})'\big)^2,
\end{eqnarray*}
where ($'$) denotes the derivative with respect to coordinate $r$. \textcolor{black}{Evaluating this expression on the solution \eqref{eq:f-N}, one finds:}
\begin{eqnarray*}
R&=&-\frac{2(z^2+3z+6)}{\ell^2}-\frac{\ell^{2(z-1)} J^2 (5-z) (3 z-19)}{2r^8}\nonumber\\
&-&\frac{3 M \ell^{z+1
}(z-1)}{r^{z+3}},
\end{eqnarray*}
leading to a curvature singularity located at the origin $r_s=0$. 
\begin{figure}[h!]
\begin{tabular}{c}
\includegraphics[width=0.48\textwidth]{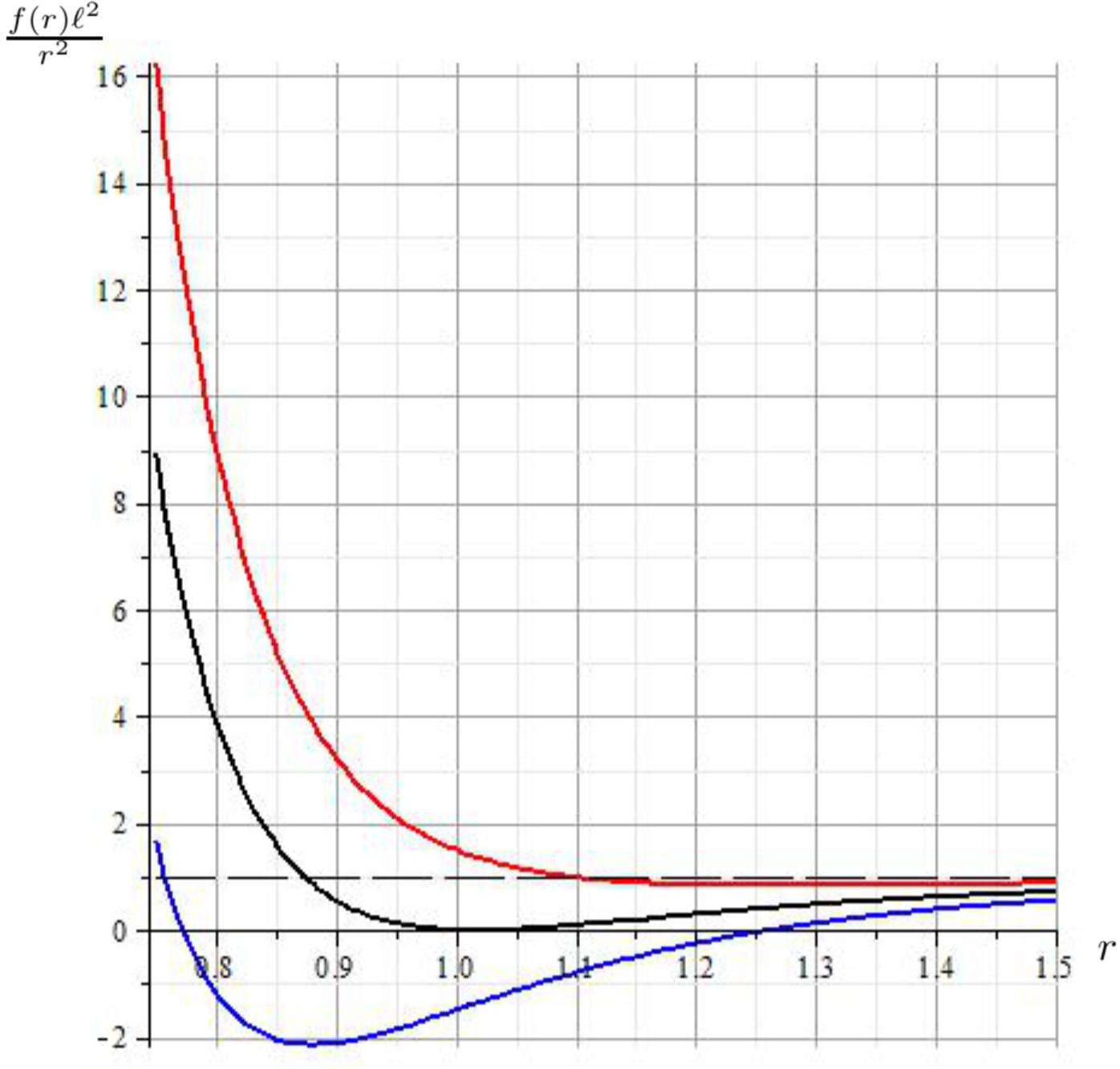}
\end{tabular}
\caption{Graphical representation of the function {${\ell^2 f(r)}/{r^{2}}$} with $\ell=1$ and $z=5/2$ for our calculations. Here, the red curve indicates the presence of a naked singularity ($f(r_{\tiny{\mbox{ext}}})>0$).  The black curve corresponds to the extremal case ($f(r_{\tiny{\mbox{ext}}})=0$). Finally, the blue curve reveals the situation $f(r_{\tiny{\mbox{ext}}})<0$, where the existence of an inner and outer horizon is ensured. }
\label{fig1}
\end{figure}

The existence of horizons is ensured, provided the following equation has a solution at $r=r_h>0$,
\textcolor{black}{$$\frac{f(r)\ell^2}{r^2}=1-\frac{M \ell^{z+3}}{r^{z+3}}+\frac{\ell^{2z} (5-z)}{2 (z-2)} \frac{J^2}{r^8}=0,$$
In the limit $r\to 0^+$, the dominant contribution is} $\frac{f(r)\ell^2}{r^2}\simeq\frac{\ell^{2z} (5-z) J^2}{2 (z-2)r^8}$, which is positive for \textcolor{black}{$z \in (2,3)$. By contrast, in the asymptotic region $r \rightarrow +\infty$,
the leading} term is $\frac{f(r)\ell^2}{r^2} \simeq1-\frac{M \ell^{z+3}}{r^{z+3}}.$ A local extreme is achieved at 
$$r_{\tiny{\mbox{ext}}}=\left[\frac{4(5-z) J^2 \ell^{z-3}}{(z+3)(z-2)M}\right]^{\frac{1}{5-z}}>0,$$
where $f''(r_{\tiny{\mbox{ext}}})>0$, and the existence of horizons is in consequence ensured if $f(r_{\tiny{\mbox{ext}}})\leq 0$ (see Figure \ref{fig1}). Notice that when the integration constants $J,M,$ vanish the spacetime \eqref{eq:metric} reduces to the static Lifshitz background.

\textcolor{black}{Building upon the analysis and framework established for this new five-dimensional charged spinning Lifshitz black hole configuration, in the next section, we will delve into the thermodynamic quantities.}


\section{Thermodynamic parameters via the Euclidean action \label{termo}}

\textcolor{black}{We now study the thermodynamics of the charged rotating Lifshitz
configuration given in eqs.~(\ref{eq:lambda})-(\ref{eq:BC}).} As demonstrated in \cite{Gibbons:1976ue,Regge:1974zd}, the partition function of a thermodynamic ensemble can be identified with the Euclidean path integral evaluated in the saddle point approximation around the Euclidean continuation of the solution, where the metric is obtained via the coordinate transformation $t=i\tau$ in the ansatz (\ref{eq:metric}) and to obtain a real metric, we can introduce a complex constant of integration for the Euclidean momentum as $J_{E}=-iJ$, where $J$ will be identified with the physical angular momentum. \textcolor{black}{The  matter} fields are given as
\begin{eqnarray*}
A^{(i)}_{\mu}dx^{\mu}=A^{(i)}_{\tau}(r)d\tau,\quad
\phi=\phi(r), \quad \psi^{(i)}=\psi^{(i)} (x_{i+1}),
\end{eqnarray*}
with $i\in \{1,2\}$. In the Euclidean continuation, the range of the radial
coordinate $r$ is from the horizon $r_h$ to infinity, and the Euclidean time is compactified as $\tau\in [0,\beta]$ where $\beta$ stands for the inverse of the \textcolor{black}{temperature $\beta=T^{-1}$, given by:
\begin{eqnarray}
T&=&\frac{N(r)f'(r)}{4\pi}\Bigg{|}_{r=r_h}\nonumber\\
&=&\frac{(z+3) r_{h}^{z}}{4 \pi \ell^{z+1}}+\frac{(5-z)^2 J^2 \ell^{z-1}}{8 \pi (z-2)  r_{h}^{8-z} }. \label{eq:Temp}
\end{eqnarray} 
With these ingredients, the Euclidean action $I_E$ can be written as:} 
\begin{equation}\label{euclideanaction}
I_E=I_{g}+I_{\phi}+I_{\psi}+B_{E},
\end{equation}
where
\begin{eqnarray}\label{euclidean-EH}
I_{g}&=&  \beta \sigma \int_{r_h}^{+\infty} dr \Big\{N \Big[\frac{3rf}{\kappa}+\frac{3r^2 f'}{2\kappa}+\frac{r^3 \Lambda}{\kappa}+\frac{\kappa p^2}{r^5}\Big]\nonumber\\
&+&N^{x_1} p'\Big\},
\end{eqnarray}
\begin{eqnarray}
I_{\phi}&=&  \beta \sigma \int_{r_h}^{+\infty} dr \Big\{N \Big[\frac{r^3 (\phi ')^2 f}{4\kappa}+\sum_{i=1}^{2}\frac{\kappa e^{-\alpha_i \phi}}{4 r^3}\mathcal{P}_{(i)}^2\Big]\nonumber\\
&+&\sum_{i=1}^{2}A^{(i)}_{\tau} \mathcal{P}_{(i)}'\Big\},\label{euclidean-phi}
\end{eqnarray}
\begin{eqnarray}
I_{\psi}&=&  \beta  \int_{r_h}^{+\infty} dr   \int_{0}^{\sigma_{1}} dx_1 \int_{0}^{\sigma_{2}} dx_2 \int_{0}^{\sigma_{3}} dx_3\,N r e^{\eta \phi}\nonumber\\
&\times&\sum_{i=1}^{2}\frac{(\partial_{x_{i+1}}\psi^{(i)})^2}{4 \kappa}.\label{euclidean-psi}
\end{eqnarray}
Here, $\partial_{x_{i+1}}\psi^{(i)}=\partial \psi^{(i)} / \partial x_{i+1}$, while that
\begin{eqnarray*}
p=\frac{r^5 (N^{x_1})'}{2N\kappa},\qquad {\cal{P}}^{(i)}= \frac{2r^3 e^{\alpha_i \phi} \left(A_{\tau}^{(i)}\right)'}{N \kappa},
\end{eqnarray*}
\textcolor{black}{correspond to the canonical momenta
associated with the rotational degree of freedom and the Abelian gauge
fields, respectively \footnote{For this situation, as before, the integration constant $\mu$ is fixed to ensure consistency with the desired asymptotic behavior, and the electric charge comes from the second vector field.}. They are directly related to the angular momentum
and the electric charge of the solution,}  and
$$\displaystyle{\sigma=\int_{0}^{\sigma_{1}} dx_1 \int_{0}^{\sigma_{2}} dx_2 \int_{0}^{\sigma_{3}} dx_3=\prod_{i=1}^{3} \sigma_i.}$$ 
\textcolor{black}{It is worth noting that the generalized Chern-Simons terms appearing in the action (\ref{eq:bgaction}) do not contribute explicitly to the Euclidean action when evaluated on the ansatz of the solution (\ref{eq:metric})-(\ref{eq:phi}). This is due to the wedge-product structure of the term $A^{(i)}\wedge H^{(i)}\wedge K^{(i)}$, which vanishes for the present field configuration. Nevertheless, these terms play an important role in the field equations and are essential for supporting the rotating Lifshitz background},
\textcolor{black}{The thermodynamic quantities follow from the on-shell Euclidean action,
which reduces to boundary contributions once the equations of motion
are imposed. In particular, the boundary term $B_E$ plays a crucial role
in ensuring a well-defined variational principle ($\delta I_E=0$) and encodes the
conserved charges associated with the solution. One may verify that the field equations obtained by varying the reduced
action (\ref{euclideanaction}) with respect to} $N, F, p ,\phi,{\cal P}^{(i)}, A_{\tau}^{(i)}, \partial_{x_{i+1}}\psi^{(i)}$ are
consistent with the rotating solution (\ref{eq:lambda})-(\ref{eq:BC}).

\textcolor{black}{To ensure a well-defined variational principle, namely $\delta I_E=0$,
one must determine the boundary term $B_E$, which encodes the
thermodynamic properties of the solution.} For this situation
$$\delta I_E=\delta I_g+\delta I_{\phi}+\delta I_{\psi}+\delta B_E=0,$$
where
\begin{eqnarray*}
\delta I_g&=&\beta \sigma
\left[N\left(\frac{3 r^2}{2 \kappa} \delta
f+\frac{r^3 f \phi^{\prime}}{2\kappa} \delta\phi\right) +\sum_{i=1}^2 A^{(i)}_{\tau}\delta{\cal
P}^{(i)}\right]_{r_h}^{+\infty},\\
\delta I_{\phi}&=&\beta \sigma
\left[N \left(\frac{3 r^2 f \phi' }{2 \kappa}\right) \delta
\phi\right]_{r_h}^{+\infty},
\\
\delta I_{\psi}&=&\beta  \int_{r_h}^{+\infty} N r e^{\eta \phi} dr \int_{0}^{\sigma_{1}} dx_1 \nonumber\\
&\times&\left\{\left[ \int_{0}^{\sigma_{3}} dx_3\,\left(\frac{\partial_{x_{2}}\psi^{(1)}}{2 \kappa}\right) \delta \psi^{(1)}\right]_{0}^{\sigma_2}\right.\nonumber\\
&+&\left.\left[ \int_{0}^{\sigma_{2}} dx_2\,\left(\frac{\partial_{x_{3}}\psi^{(2)}}{2 \kappa}\right) \delta \psi^{(2)}\right]_{0}^{\sigma_3}\right\}.
\end{eqnarray*}
The variations of the 
boundary term at infinity, as well as at the horizon, take the form:
\begin{eqnarray}
\delta B_E\Big{|}_{+\infty}&=&\beta \delta \left(\frac{3 \ell^2 M  \sigma }{2 \kappa}\right)\\
&-&\beta\Phi_e \delta \left(\frac{\sqrt{3(3-z)}\,(5-z) J \ell^{z-1} \mu^{\frac{1}{\sqrt{6(z-1)}}}\sigma }{ \sqrt{z-2}\,\kappa }\right),\nonumber\\
\delta B_E\Big{|}_{r_h}&=&\beta \Omega \delta \left( -\frac{(5-z) J \ell^{z-1} \sigma}{2 \kappa}\right)\nonumber\\
&+&{ \beta \sum_{i=1}^{2} \Psi^{(i)}_{a}\delta \left(\frac{(5-z)^2 \ell^{2(z-1)} J \mu^{\frac{6}{\sqrt{6(z-1)}}} \sigma}{ 2 \kappa}\right)}\nonumber\\
&+& \delta\left(\frac{2 \pi r_h^3 \sigma}{\kappa}\right),\label{eq:var}
\end{eqnarray}
with \textcolor{black}{the electric potential $\Phi_e$ and angular velocity $\Omega$ given by
\begin{eqnarray}
\Omega&=& -\frac{J}{r_{h}^{5-z}}\,,\quad 
\Phi_e= \sqrt{\frac{3(3-z)}{4(z-2)}}\frac{J}{ \mu^{\frac{1}{\sqrt{6(z-1)}}} r_h^{5-z}}, \label{eq:Omega}
\end{eqnarray}
while that of the axionic potential $\Psi_a$ is identified as
\begin{eqnarray}
\Psi^{(1)}_a&=&\Psi^{(2)}_a=\int_{r_h}^{+\infty} (J N(r) r e^{\eta \phi(r)})dr \nonumber\\
&=&\Psi^{(2)}_a= \frac{J}{(5-z) \ell^{z-1}\mu^{\frac{6}{\sqrt{6(z-1)}}} r_h^{5-z}}.\label{eq:psia} 
\end{eqnarray}
Finally,} the boundary term $B_E$ can be expressed as

\begin{eqnarray*}
B_E &=&\beta  \left(\frac{3 \ell^2 M  \sigma }{2 \kappa}\right)\nonumber\\
&-&\beta\Phi_e \left(\frac{\sqrt{3(3-z)}\,(5-z) J \ell^{z-1} \mu^{\frac{1}{\sqrt{6(z-1)}}}\sigma }{ \sqrt{z-2}\,\kappa }\right),
\end{eqnarray*}
\begin{eqnarray}
&-&\beta \Omega  \left( -\frac{(5-z) J \ell^{z-1} \sigma}{2 \kappa}\right)\label{eq:Be}\\
&-&{\beta \sum_{i=1}^{2}\Psi^{(i)}_{a} \left(\frac{(5-z)^2 \ell^{2(z-1)} J \mu^{\frac{6}{\sqrt{6(z-1)}}} \sigma}{2 \kappa}\right)}\nonumber\\
&-&\left(\frac{2 \pi r_h^3 \sigma}{\kappa}\right).\nonumber
\end{eqnarray}
The Euclidean action
on-shell reduces to the boundary term and is related to the Gibbs free energy ${\cal G}$ as
\begin{eqnarray}
I_E\Big{|}_{\tiny{\mbox{on-shell}}}&=&B_E=\beta {\cal G}=\beta\Big({\cal M}-\Phi_e{\cal Q}_e-\Omega {\cal J}\nonumber\\
&-&{\sum_{i=1}^2\Psi^{(i)}_a \omega^{(i)}_a} \Big)-{\cal S}, \label{eq:Gibbs}
\end{eqnarray}
where ${\cal M}$ is the mass, ${\cal S}$ the entropy, \textcolor{black}{and ${\cal Q}_e$ represents the
electric charge, while, as before,} $\Phi_e$ corresponds to the electric potential, $\Omega$ and the $\Psi^{(i)}_a$ are the chemical potentials corresponding to the angular momentum ${\cal J}$ and the axionic charges $\omega^{(i)}_a$, \textcolor{black}{respectively}. Via the comparison between eqs. (\ref{eq:Be}) and (\ref{eq:Gibbs}), the thermodynamics quantities \textcolor{black}{take the form:}
\begin{eqnarray}
\label{Lifqtes}
\mathcal{M}&=&\frac{3 \ell^2 M  \sigma }{2 \kappa}=\frac{3 r_{h}^{z+3} \sigma}{2 \kappa \ell^{z+1}}+\frac{3 (5-z) J^2 \ell ^{z-1} \sigma}{4 \kappa (z-2) r_{h}^{5-z} }\,,\nonumber\\
\mathcal{S}&=&\frac{2 \pi r_h^3 \sigma}{\kappa}\,,\quad 
\mathcal{J}=-\frac{(5-z) J \ell^{z-1} \sigma}{2 \kappa}\,,\nonumber\\
\mathcal{Q}_e&=&\frac{\sqrt{3(3-z)}\,(5-z) J \ell^{z-1} \mu^{\frac{1}{\sqrt{6(z-1)}}}\sigma }{ \sqrt{z-2}\,\kappa }\,,\nonumber\\
{\omega_a^{(1)}}&=&\omega_a^{(2)}=\frac{(5-z)^2 \ell^{2(z-1)} J \mu^{\frac{6}{\sqrt{6(z-1)}}} \sigma}{ 2\kappa},\label{eq:wa} 
\end{eqnarray}
while the \textcolor{black}{Hawking temperature is given in (\ref{eq:Temp}) and the} corresponding conjugate, intensive quantities \textcolor{black}{are given in eqs. (\ref{eq:Omega})-(\ref{eq:psia}).}

\textcolor{black}{The thermodynamical quantities (\ref{eq:wa}) are characterized by the horizon radius $r_h$ and the constant $J$. Varying these parameters induces variations
of the thermodynamic quantities as follows:
\begin{eqnarray*}
\delta{\cal M}&=& \left(\frac{3 (z+3) r_{h}^{z+2} \sigma}{2 \kappa \ell^{z+1}}-\frac{3 (5-z)^2 J^2 \ell ^{z-1} \sigma}{4 \kappa (z-2) r_{h}^{6-z} }\right)\delta r_h \\
&+& \left(\frac{3 (5-z) J \ell ^{z-1} \sigma}{2 \kappa (z-2) r_{h}^{5-z} }\right) \delta J,\\
\delta{\cal S}&=& \left(\frac{6 \pi r_h^2 \sigma}{\kappa}\right) \delta r_h, \\
\delta{\cal Q}_{e}&=& \left(\frac{\sqrt{3(3-z)}\,(5-z)  \ell^{z-1} \mu^{\frac{1}{\sqrt{6(z-1)}}}\sigma }{ \sqrt{z-2}\,\kappa }\right) \delta J,\\
\delta {\cal J}&=& \left(-\frac{(5-z)  \ell^{z-1} \sigma}{2 \kappa}\right) \delta J,
\end{eqnarray*}
\begin{eqnarray*}
\delta{\omega}^{(i)}_a&=& \left( \frac{(5-z)^2 \ell^{2(z-1)}  \mu^{\frac{6}{\sqrt{6(z-1)}}} \sigma}{ 2\kappa}\right) \delta J.
\end{eqnarray*}
Using the variations displayed above, one finds that the thermodynamic
quantities satisfy the first law of black-hole thermodynamics:}
\begin{eqnarray}\label{eq:first-law}
\delta{\cal M}=T \delta{\cal S}+\Phi_e \delta{\cal Q}_{e}+\Omega \delta{\cal J}+{\sum_{i=1}^{2}\Psi^{(i)}_a \delta{\omega}^{(i)}_a.}
\end{eqnarray}
\textcolor{black}{Moreover, the mass ${\cal M}$ can be regarded as a homogeneous function
of the extensive variables ${\cal S}$, ${\cal J}$, ${\cal Q}_e$, and
$\omega_a^{(i)}$, once their scaling behavior under rescalings of
$r_h$ and $J$ is taken into account. This allows us to apply Euler's theorem for homogeneous functions to derive the corresponding Smarr relation. Under the rescalings:} 
\begin{equation}
r_h\to \lambda r_h,\qquad J\to \lambda^4 J,
\end{equation}
the thermodynamic \textcolor{black}{parameters} scale as
\begin{eqnarray} 
&&{\cal M}\to \lambda^{z+3} {\cal M}\,,\ {\cal J}\to \lambda^4 {\cal J}\,,\ {\cal Q}_e\to \lambda^4 {\cal Q}_e\,,\nonumber\\
&&\ {\cal S}\to \lambda^3 {\cal S},\,{\omega^{(i)}_a\to \lambda^{4} \omega^{(i)}_a\,.}
\end{eqnarray}
By applying the Euler scaling argument for homogeneous functions, one obtains the following Smarr formula:
\begin{equation}
{\cal M}=\left(\frac{1}{z+3}\right) \left(3 T{\cal S}+4 \Omega{\cal J}+ {4\sum_{i=1}^{2}\Psi^{(i)}_a {\omega}^{(i)}_a}\right).\label{eq:smarr}
\end{equation}
For $z \neq 3$, the following relation can be obtained
\begin{equation}
\Omega{\cal J}=\frac{(z-2)}{3(3-z)} \Phi_e {\cal Q}_e={\Psi^{(i)}_a \omega^{(i)}_a,}\label{eq:relation}
\end{equation}
allowing us to simplify the Smarr
relation (\ref{eq:smarr}), for example, in the following structure:
\begin{equation}
{\cal M}=\left(\frac{1}{z+3}\right) \left[3 T{\cal S}+\left(\frac{12}{z-2}\right) \Omega{\cal J}\right].\label{eq:smarr2}
\end{equation}

\section{Holographic Superconductor}\label{hol-sup}
In the previous section we have constructed the first exact, rotating, Lifshitz black hole, and we have understood its thermal properties. Now, we investigate the holographic implications of the preceding black hole configuration by focusing on the construction of a s-wave holographic superconductor. \textcolor{black}{Following the standard approach to holographic superconductors, we
introduce an additional matter sector} to \eqref{eq:bgaction} composed of a Maxwell field coupled to a complex scalar field, described by the action \cite{Gubser:2008px, Hartnoll:2008vx, Hartnoll:2008kx}:
\begin{equation}\label{eq:matter_sup}
	S_{\text{M}} = -\frac{1}{q^2} \int \textcolor{black}{d^5x} \sqrt{-g} \Big( \frac{1}{4} F_{\mu\nu} F^{\mu\nu} + \frac{1}{\ell^2} \big( |D_\mu \Psi|^2 + m^2 |\Psi|^2 \big) \Big),
\end{equation}
where \( F_{\mu\nu} = 2\partial_{[\mu} A_{\nu]} \) is the field strength tensor, and \( \Psi \) is a complex scalar field of mass \( m \) and charge \( q \), minimally coupled to the gauge field \( A_\mu \) via the covariant derivative \( D_\mu = \nabla_\mu - i A_\mu \). {{Notice that we can rescale $A_\mu\rightarrow q A_\mu$ and $\Psi\rightarrow \ell q\Psi$, to recover the standard action of scalar QED.}}

\textcolor{black}{As is standard in the study of holographic superconductors, we work in} the \textit{probe limit} \( q \gg 1 \), where the matter fields do not backreact on the background geometry. In this limit, the gravitational sector decouples, allowing us to treat the background as fixed and focus solely on the dynamics of the gauge and scalar fields.

The resulting equations of motion derived from the action \eqref{eq:matter_sup} are:
\begin{subequations}\label{eq:GenFieldEqns_sup}
	\begin{align}
		\nabla_\mu F^{\mu\nu} - \frac{1}{\ell^2} \left( 2 A^\nu |\Psi|^2 - i \Psi \nabla^\nu \overline{\Psi} + i \overline{\Psi} \nabla^\nu \Psi \right) &= 0, \label{eq:gen_max_sup} \\
		\nabla^\mu \nabla_\mu \Psi - 2i A^\mu \nabla_\mu \Psi - i \Psi \nabla_\mu A^\mu - (m^2 + A_\mu A^\mu) \Psi &= 0. \label{eq:gen_sca_sup}
	\end{align}
\end{subequations}
These equations are to be solved on the fixed gravitational background \eqref{eq:metric}, subject to appropriate boundary conditions.

To facilitate our analysis, we express the metric function $f(r)$ in terms a new function $h(r)$ defined by $f(r) = (r/\ell)^2 h(r)$, where $h(r)$ satisfies the boundary conditions 
$$\lim_{r\to r_h}h(r) = 0\, \mbox{ and} 
\lim_{r\to +\infty}h(r) = 1.$$ 
In addition, we redefine the holographic coordinate as $u=r_h/r$, such that the horizon {as well as} the asymptotic boundary  are located at $u=1$ and $u=0$, respectively. 

\subsection{Condensation of the scalar field}
To explore the condensation of the order parameter, we consider a real scalar field and a Maxwell field with the simplest coordinate {dependence:} 
\begin{equation}\label{eq:ansatz_condensate}
	\Psi=\Psi(u), \qquad {A=A_t(u) dt + A_{x_1}(u) dx_1}.
\end{equation}
\textcolor{black}{With this ansatz, the system of equations \eqref{eq:gen_max_sup}-\eqref{eq:gen_sca_sup} take} the form
\begin{subequations}\label{eq:CondeFieldEqns}
\begin{align}
\Psi'' &+ \Big(\dfrac{h'}{h} -\dfrac{z+2}{u} \Big)\Psi'-\dfrac{\ell^2}{u^2 r_h^2 h}\Big[ m^2r_h^2+ u^2 A_{x_1}^2 \nonumber\\ 
&- \dfrac{\ell^{2z} r_h^{2-2z} u^{2z}}{h} \left( A_t- N^{x_1}A_{x_1} \right)^2 \Big] \Psi = 0,\\
A_t'' &+ \Big[\dfrac{z-2}{u} +  \ell^{2z} \Big(\dfrac{u}{r_h}\Big)^{2z-2}\dfrac{(N^{x_1})' N^{x_1}}{h} \Big] A_t' \nonumber \\ &- \Big[\Big(h+ \ell^{2z}\Big(\dfrac{u}{ r_h}\Big)^{2z-2} {(N^{x_1})}^2\Big)\dfrac{(N^{x_1})'}{h} - \Big( \dfrac{h'}{h} \nonumber \\& - \dfrac{2(z-1)}{u} \Big)N^{x_1} \Big] A_{x_1}' - \dfrac{2\psi^2}{u^2 h} A_t = 0,
\end{align}
\begin{align}
A_{x_1}'' &+ \Big[\dfrac{h'}{h}-\dfrac{z}{u} -  \ell^{2z} \Big(\dfrac{u}{r_h}\Big)^{2z-2}\dfrac{(N^{x_1})' N^{x_1}}{h} \Big] A_{x_1}'\nonumber \\ &+ \ell^{2z}\Big(\dfrac{u}{r_h}\Big)^{2z-2}\dfrac{(N^{x_1})'}{h} A_{t}'-\dfrac{2\psi^2}{u^2 h} A_{x_1} = 0,
\end{align}
\end{subequations}
\textcolor{black}{where primes now denote derivatives with respect to the radial coordinate $u$.}

To solve this system, we must impose suitable boundary conditions at both the horizon (\( u = 1 \)) and the conformal boundary (\( u = 0 \)) of the spacetime. 
At the horizon, regularity requires the scalar field to remain finite, \( \Psi(1) < \infty \), and the temporal and angular components of the gauge field to vanish, \( A_t(1) = 0 \) and \( A_{x_1}(1) = 0 \), ensuring that \( A_\mu \) has finite norm within the bulk.
Near the boundary, the fields asymptotically behave as 
\begin{align}
\Psi &= \Psi_{1}u^{\Delta_{1}}+\Psi_{2}u^{\Delta_{2}},\label{eq:BoundarySols1}\\ A_t&=\mu_{t}-\rho\Big(\dfrac{u}{r_h}\Big)^{3-z},\quad A_{x_1}=\mu_{x_1}-J_{x_1} \Big(\dfrac{u}{r_h}\Big)^{z+1},\label{eq:BoundarySols2}
\end{align}
with scaling dimensions
\begin{equation}\label{eq:scaling_dim}
	\Delta_{1,2}= \dfrac{1}{2}\left[(z+3)\pm \sqrt{(z+3)^2+4 m^2\ell^2}\right].
\end{equation}
\textcolor{black}{Notice that this $A_t$ is valid for $z\neq3$. At $z=3$ logarithmic
terms appear in the asymptotic behavior of the gauge field, which would
require a separate analysis and will
not be considered here.}
The coefficient $\Psi_1$ ($\Psi_2$) corresponds to the order parameter $\mathcal{O}$ with scaling dimension $\Delta_1$ ($\Delta_2$). To ensure the existence of well-defined normalizable modes near the boundary regime, the mass of the scalar field must satisfy the bound $m^2\ell^2 \ge -(z+3)^2/4$, which is the Breitenlohner-Freedman (BF) bound for massive scalar fields in a 5-dimensional Lifshitz background. In addition, the parameters $\mu_{t}$ and $\rho$ are interpreted as the chemical potential and the charge density operator, respectively. A similar interpretation is given to the coefficients of the component $A_{x_1}$, where $\mu_{x_1}$ represents a one-dimensional vector potential that sources the  current density operator $J_{x_1}$.

\textcolor{black}{To determine the condensate as a function of temperature, we solve the
system \eqref{eq:CondeFieldEqns} numerically subject to the boundary
conditions \eqref{eq:BoundarySols1}--\eqref{eq:BoundarySols2}. We work
in the canonical ensemble, keeping the charge density $\rho$ fixed.} \textcolor{black}{For the numerical analysis, we choose the lower scaling dimension
$\Delta_1=1$. From eq.~\eqref{eq:scaling_dim}, this fixes the scalar
mass to $
m^2\ell^2=-(z+2).$ After the rescalings $A_\mu\to qA_\mu$ and $\Psi\to \ell q\Psi$, the
effective scalar charge becomes unity. In the boundary expansion of $A_{x_1}$, we set the source
$\mu_{x_1}=0$ in order to avoid introducing an external current in the
dual field theory. We also impose the source-free condition
$\Psi_2=0$, so that $\Psi_1$ is proportional to the expectation value
of the dual operator, namely $\Psi_1\sim \langle \mathcal{O}_1\rangle$.}


\textcolor{black}{We solve the coupled ordinary differential equations \eqref{eq:CondeFieldEqns} using a shooting method from the horizon to the boundary.
At the horizon, we expand the fields compatible with regularity at $u=1$
\begin{subequations}\label{eq:NHfields}
\begin{align}
\Psi(u) &= \Psi_h + \Psi_h'(u-1) + \mathcal{O}((u-1)^2),\\
A_t(u) &= a_t^{(1)}(u-1) + \mathcal{O}((u-1)^2),\\
A_{x_1}(u) &= a_{x_1}^{(1)}(u-1) + \mathcal{O}((u-1)^2),
\end{align}
\end{subequations}
where the coefficients $\Psi_h$, $a_t^{(1)}$, and $a_{x_1}^{(1)}$ are related through the equations of motion \eqref{eq:CondeFieldEqns}, leaving two independent parameters (in our case, $\Psi_h$ and $a_t^{(1)}$).}

\begin{figure*}[htbp]
	\centering
\includegraphics[width=\columnwidth]{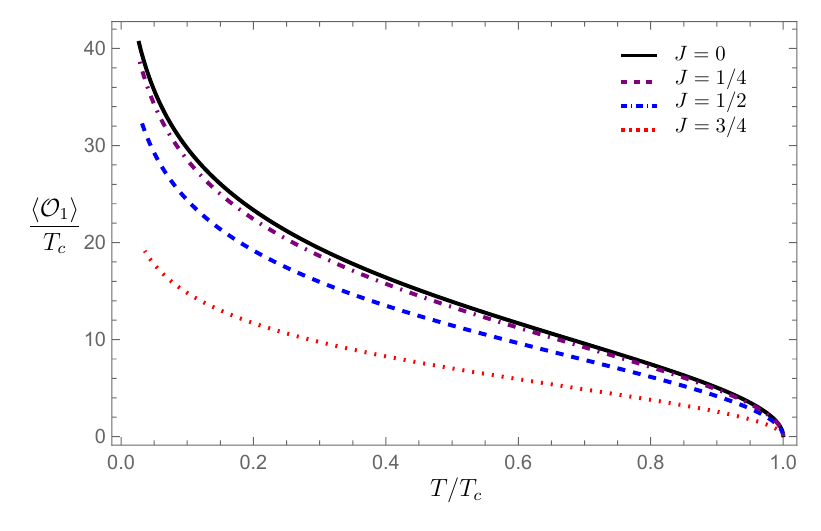}
\includegraphics[width=\columnwidth]{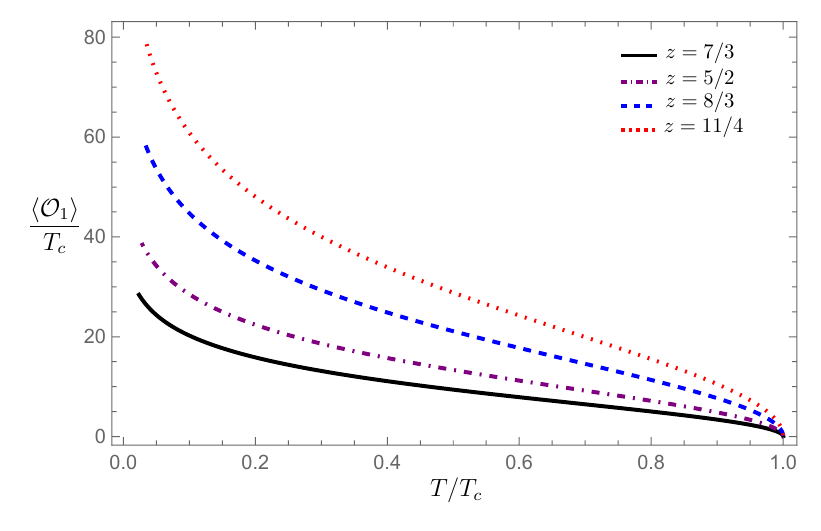}
	\caption{The condensation profiles for the operator $\mathcal{O}_1$ as function of temperature, considering different values of the rotation parameter $J$ (left panel) and the dynamical exponent $z$ (right panel). 
    }
	\label{fig:Condensate}
\end{figure*}

\textcolor{black}{Starting from the near-horizon expansions \eqref{eq:NHfields}, the equations are integrated
toward the boundary at $u=0$. The shooting parameters $(\Psi_h, a_t^{(1)})$ are adjusted
such that for each temperature (or equivalently $r_h$) the source-free condition $\Psi_2=0$ is satisfied. Once the boundary conditions are met, the coefficient $\Psi_1 \sim \langle \mathcal{O}\rangle$ is extracted as function of the temperature in order to construct the condensation curves. The critical temperature $T_c$ is determined by the onset of a non-trivial solution, or the temperature at which $\langle \mathcal{O}\rangle$ vanishes.  Our computations are performed setting $\ell=1$, while varying the rotation parameter $J$ at fixed critical exponent $z$, and by varying $z$ while keeping $J$ fixed.
In addition, we verified that the numerical solutions are stable when changing the integration step and the initial guesses for the shooting parameters.} The results are presented in Figure~\ref{fig:Condensate}. Specifically, the left panel illustrates the effect of increasing the rotation parameter, considering the values $J=0$, $1/4$, $1/2$ and $3/4$ while keeping the critical exponent fixed at $z = 5/2$. As shown, increasing the rotation leads to a reduction in the amplitude of the condensate. On the other hand, the right panel shows the influence of varying the critical exponent, with $z=7/3$, $5/2$, $8/3$ and $11/4$ for a constant value of the rotation parameter $J = 1/4$. Contrary to the effect of rotation, increasing the critical exponent results in an enhancement in the amplitude of the condensate. Altogether, these results suggest that the anisotropy induced by a larger critical exponent favors the formation of the superconducting phase.

\subsection{The AC conductivity}
\begin{figure*}[htbp]
	\centering
\includegraphics[width=\columnwidth]{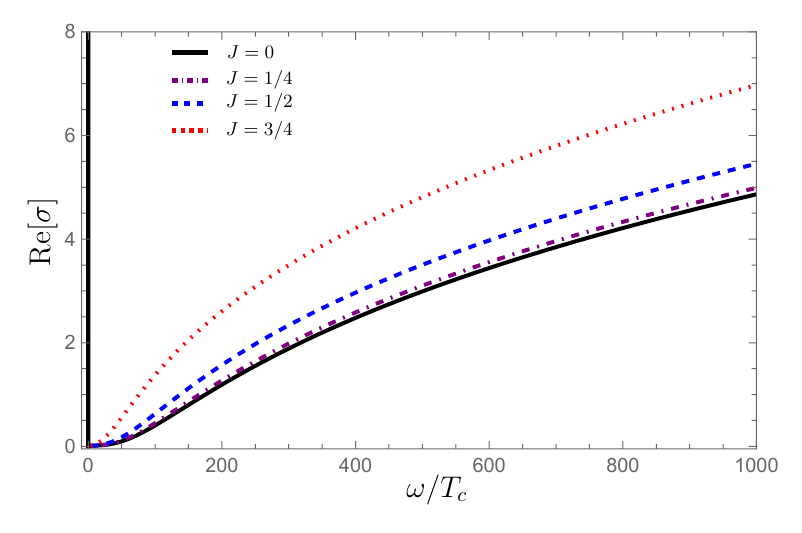}
\includegraphics[width=\columnwidth]{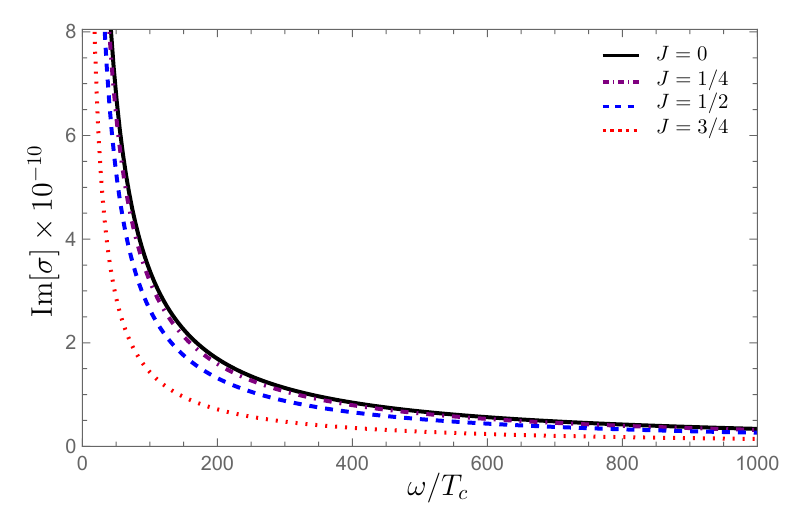}
	\caption{Real (left) and imaginary (right) parts of the conductivity as functions of the frequency for different values of the rotation parameter. These plots were generated at $T\approx 0.036\,T_c$, considering $\Delta =1$, $z=5/2$ and $\ell =1$. 
    }
\label{fig:sigma1}
\end{figure*}
\begin{figure*}[htbp]
	\centering
\includegraphics[width=\columnwidth]{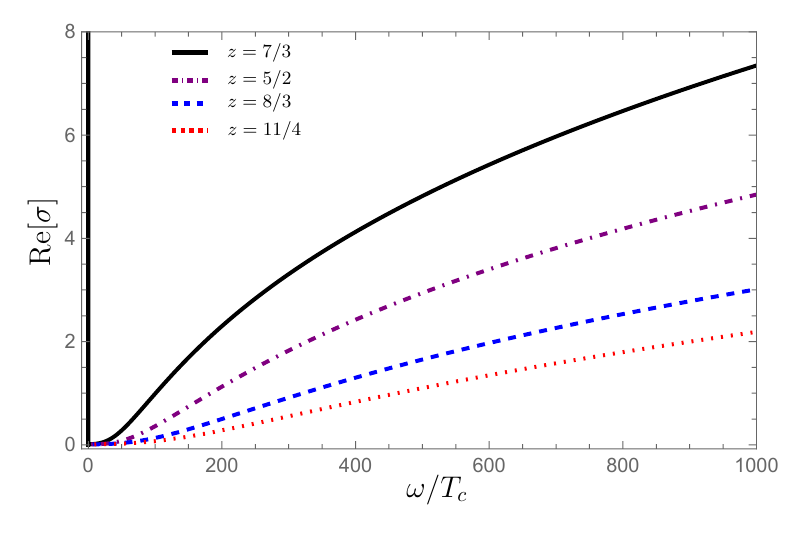}
\includegraphics[width=\columnwidth]{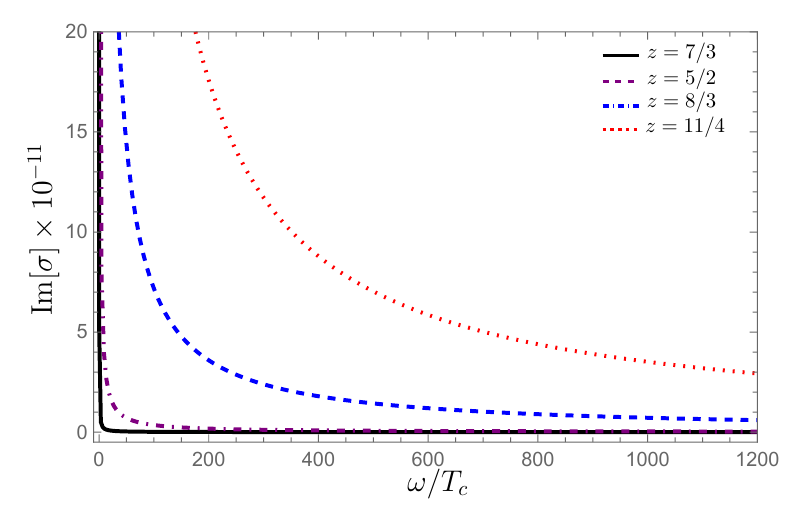}
	\caption{Real (left) and imaginary (right) parts of the conductivity as functions of the frequency for different values of the dynamical critical exponent. These curves were generated at $T\approx 0.036\,T_c$, considering $\Delta =1$, $J=1/4$ and $\ell =1$. 
    }
\label{fig:sigma2}
\end{figure*}

The AC conductivity is a fundamental quantity in the description of a superconductor, as it represents the response of the system to external electric perturbations. In the holographic framework, it is computed by introducing a gauge field perturbation in the bulk and solving the resulting linearized Maxwell equation. To this end, we consider a harmonic perturbation along the $x_2$-component of the gauge field, given by $\delta A_{x_2} = A_{x_2}(u) e^{-i \omega t} dx_2$. Substituting this ansatz into the system \eqref{eq:GenFieldEqns_sup}, we obtain
\begin{equation}\label{eq:maxeq_Ax}
	A_{x_2}''+\Big(\frac{h'}{h}-\frac{z}{u}\Big)A_{x_2}'+\Biggl[\Big(\frac{\ell u}{r_{h}}\Big)^{2z}\frac{\omega^2\ell^2}{h} - 2 \Psi^2 \Biggr]\frac{A_{x_2}}{u^2 h} =0,
\end{equation}
which, in order to solve it, we impose the ingoing wave condition at the horizon
\begin{equation}\label{eq:AxH}
	A_{x_2} = (u-1)^{-\frac{i \omega 
    }{4\pi T}}\Big[1+A_{01}(u-1)+A_{02}(u-1)^2+\cdots\Big],
\end{equation}
while the asymptotic boundary behavior is given by
\begin{equation}\label{eq:AxB}
A_{x_2} = A_{x_2}^{(0)} + A_{x_2}^{(z+1)}\Big(\frac{u}{r_{h}}\Big)^{z+1}+\cdots,
\end{equation}
where the coefficient $A_{x_2}^{(0)}$ is interpreted as the source of a conjugated current $J_{x_2}$, with $A_{x_2}^{(z+1)}$ related to the expectation value, $\langle J_{x_2}\rangle \sim A_{x_2}^{(z+1)}$. Applying the Ohm's law, we obtain an expression for the AC conductivity 
\begin{equation}\label{eq:Ohms_law}
	\sigma(\omega) = \frac{\langle J_{x_2}\rangle}{E_{x_2}} = -\dfrac{(z+1)i}{\omega}\dfrac{ A_{x_2}^{(z+1)}}{ A_{x_2}^{(0)}}.
\end{equation}
\textcolor{black}{Although the fluctuation equation \eqref{eq:maxeq_Ax} is linear once the
background is fixed, obtaining an analytic solution is still difficult.
We therefore solve it numerically by integrating from the horizon to the boundary, imposing ingoing wave conditions at the horizon \eqref{eq:AxH} and extracting the conductivity from the asymptotic behavior of the solution \eqref{eq:AxB}.}

Figure~\ref{fig:sigma1} illustrates the influence of the background rotation on the AC conductivity as a function of frequency while keeping the critical exponent fixed. \textcolor{black}{As $\omega/T_c \to 0$,
$\mathrm{Re}[\sigma]$ tends to vanish, 
while $\mathrm{Im}[\sigma]$
shows a divergent behavior in this regime which indicates the existence of a pole at $\omega=0$. According to the Kramers--Kronig relations,
\begin{equation}
    \operatorname{Re}[\sigma(\omega)] = \frac{1}{\pi} \mathcal{P} \int_{-\infty}^{\infty} \frac{\operatorname{Im}[\sigma(\omega')]}{\omega' - \omega} d\omega',
\end{equation}
such a pole at $\omega=0$ implies a
delta-function contribution in the real part, $\mathrm{Re}[\sigma] \sim \pi \delta[\omega]$ at $\omega =0$. Such a feature is the hallmark of perfect DC conductivity that signals the onset of the superconducting phase \cite{Hartnoll:2008vx}.} In addition, this low-frequency suppression of $\text{Re}[\sigma]$ signals the opening of a frequency gap, which becomes less pronounced as the rotation parameter increases, suggesting that the superconducting state is weakened by the presence of the rotation. At high frequencies, $\text{Re}[\sigma]$  shows a smooth, monotonic increase, while $\text{Im}[\sigma]$ gradually decays to zero, indicating a transition to a normal state.

In contrast, in Figure~\ref{fig:sigma2} we examine the impact of the critical exponent on the conductivity while keeping the rotation parameter fixed. A qualitatively similar behavior is observed here compared to the preceding situation. However, in this case, increasing the degree of anisotropy---encoded in the critical exponent---enhances the superconducting state, as indicated by a sharper frequency gap and a more pronounced suppression of $\text{Re}[\sigma]$ at low frequencies.

\section{Conclusions and discussions }\label{conclusions}

In the present work, we construct a novel family of exact five-dimensional rotating asymptotically Lifshitz black holes carrying electric and axionic charges, supported by a dilaton, Abelian gauge fields, axionic scalars, and two generalized Chern-Simons terms. These spacetimes describe black hole configurations with well-defined horizons shielding a curvature singularity, and we have determined the range of the dynamical critical exponent $z$ for which real solutions exist. As shown in Figure \ref{fig1}, the analysis of the metric function $f$ reveals the presence of inner and outer horizons, \textcolor{black}{naked singularities}, and extremal configurations, allowing us to obtain a clear characterization of the horizon structure of the solutions. \textcolor{black}{The construction of explicit rotating Lifshitz geometries is known to be technically challenging, and exact solutions in this context
remain relatively scarce. The solutions presented here, therefore, provide a new gravitational background that extends the class of known Lifshitz black holes and offers a useful framework for exploring non-relativistic holography beyond static configurations.}

{With respect to the study of the thermodynamic properties of these black holes, we compute the conserved charges and their associated chemical potentials, showing that the solutions satisfy a consistent first law of black hole thermodynamics (\ref{eq:first-law}). Furthermore, by exploiting the scaling properties of the system, we derived a Smarr relation that explicitly captures the role played by anisotropic scaling, rotation, and the additional matter fields (\ref{eq:smarr}). In fact, through the relation (\ref{eq:relation}), the Smarr formula can be re-expressed by eq. (\ref{eq:smarr2}).  The presence of axionic fields plays a crucial role in supporting the rotating Lifshitz geometry and in ensuring the consistency of the conserved charges and thermodynamic relations. }

\textcolor{black}{As an application of this gravitational background}, we analyzed the condensation of the order parameter and the electric response of the system using the numerical shooting method. A goal of this work was to study the effects of the black hole rotation on the essential properties of the holographic superconductor model. 
Our findings show that, when considering a fixed value of the critical exponent $z$, increasing the black hole rotation reduces the amplitude of the condensate. This is consistent with the behavior depicted in Fig. \ref{fig:sigma1}, where the dissipative effects, contained in $\text{Re}[\sigma]$, are more significant when the rotation of the background is increased. In this scenario, the frequency gap is decreased, and the system loses the ability to superconduct. Consequently, the superconducting response of the system, encoded in $\text{Im}[\sigma]$, is suppressed with the rotation. Conversely, when keeping the rotation parameter fixed and varying the dynamical critical exponent $z$, we observed that an increase in $z$ leads to an increment in \textcolor{black}{ the condensate amplitude}. This behavior aligns with the results given in Fig. \ref{fig:sigma2}, which shows a reduction in the dissipative effects and a corresponding amplification of the superconducting response as the critical exponent is increased. Unlike the previous scenario, the width of the gap now enlarges, indicating a strengthening of the superconducting phase. Altogether, these results suggest that, \textcolor{black}{in this holographic superconductor model}, the anisotropy of the background favors the formation of the superconducting state whereas the rotation acts as a competing effect that tends to suppress it.

It should be noted that the effects induced by the rotation here contrast with those reported in \cite{Srivastav:2019ixc, Herrera-Mendoza:2024vfj}, where the presence of the rotation was found to favor the superconducting phase. \textcolor{black}{This discrepancy likely arises from the fact that} the models considered in these papers were constructed on backgrounds that are not asymptotically AdS or Lifshitz; in such setups, recovering an AdS or Lifshitz structure requires turning off the rotation parameter. In contrast, the model considered here is built on a rotating Lifshitz background that behaves asymptotically Lifshitz when the rotation parameter is turned off. Therefore, the difference in the asymptotic structure is responsible for the appearance of the novel behavior reported in this manuscript.

Finally, this study opens several new directions for future research, including the analysis of dynamical and thermodynamic stability, the construction of fully backreacted matter configurations, the exploration of higher-dimensional generalizations, {through a Kerr-Schild transformation, as was performed in \cite{Ortaggio:2023rzp}}, and the investigation of rotating Lifshitz black holes as backgrounds for non-relativistic holographic hydrodynamics. We expect that the new family of solutions presented here will provide a useful laboratory for further developments in the study of anisotropic and rotating spacetimes.

\acknowledgments
\textcolor{black}{The authors would like to thank the anonymous referee for carefully reading our
manuscript and giving valuable suggestions that led to an improved version of this work.} MB is supported by Proyecto Interno  UCM-IN-25202 l\'inea regular. XZ is supported by National Natural Science Foundation of China (NSFC) with
Grants No.12275087. JAHM is grateful
to SNII from SECIHTI. This work is also partially funded by FONDECYT grants 1262452 and 1250133.


\end{document}